\documentclass[sigconf]{acmart}
\pdfoutput=1
\usepackage{xcolor,graphicx,array,pifont,listings}
\newcolumntype{L}{>{\centering\arraybackslash}m{3cm}}

\usepackage{acronym}
\acrodef{LSM}{Linux Security Module}
\copyrightyear{2024}
\acmYear{2024}
\setcopyright{rightsretained}
\acmConference[SESoS '24 ]{2024 12th ACM/IEEE International Workshop on Software Engineering for Systems-of-Systems and Software Ecosystems}{April 14, 2024}{Lisbon, Portugal}
\acmBooktitle{2024 12th ACM/IEEE International Workshop on Software Engineering for Systems-of-Systems and Software Ecosystems (SESoS '24 ), April 14, 2024, Lisbon, Portugal}\acmDOI{10.1145/3643655.3643878}
\acmISBN{979-8-4007-0557-1/24/04}

\begin{document}

\title{Sandboxing Adoption in Open Source Ecosystems}
\author{Maysara Alhindi}
\affiliation{\institution{University of Bristol} \country{United Kingdom}}
\email{maysara.alhindi@bristol.ac.uk}
\author{Joseph Hallett}
\affiliation{\institution{University of Bristol} \country{United Kingdom}}
\email{joseph.hallett@bristol.ac.uk}

\begin{abstract}

Sandboxing mechanisms allow developers to limit how much access applications have to resources, following the least-privilege principle. However, it's not clear how much and in what ways developers are using these mechanisms. This study looks at the use of \emph{Seccomp}, \emph{Landlock}, \emph{Capsicum}, \emph{Pledge}, and \emph{Unveil} in all packages of four open-source operating systems. We found that less than 1\% of packages directly use these mechanisms, but many more indirectly use them. Examining how developers apply these mechanisms reveals interesting usage patterns, such as cases where developers simplify their sandbox implementation. It also highlights challenges that may be hindering the widespread adoption of sandboxing mechanisms.

\end{abstract}

\maketitle
\section{Introduction}

You can't always trust that the software you use will behave as expected. Attackers may exploit privileged software to increase their level of access. Sandboxes help prevent this by following the \emph{principle of the least privilege}~\cite{Saltzer_Schroeder_1975}, ensuring that programs run with the minimum functionality needed. If a program shouldn't access specific files or network devices, the sandbox prevents it, safeguarding against potential exploits.

In open-source ecosystems, sandboxes play a crucial role as systems are constructed by combining software into \emph{packages} to meet specific deployment requirements. While most contemporary operating systems offer some type of sandboxing mechanisms, unlike many OS features, sandboxing functionality (aside from basic \texttt{chroot}s and the Unix DAC) lacks standardization. Each operating system presents unique mechanisms, at times featuring multiple competing options. Consequently, if software intends to leverage sandboxing, it needs customization to align with each OS it aims to support.

To ensure software is properly tailored for an operating system, open source OS vendors provide package management tools and repositories. In these repositories, software is adjusted to fit within the OS, disabling unsupported features, aligning installation paths, and applying patches for OS-specific functionality.

Various operating systems offer different sandboxing mechanisms. Linux employs \emph{Seccomp} and the \ac{LSM} framework, allowing for pluggable kernel modules like \emph{Landlock}. FreeBSD has \emph{Capsicum} for sandboxing, while OpenBSD enables developers to sandbox applications using \emph{Pledge} and \emph{Unveil} by specifying system calls and file paths.

To what degree do software packages in various open-source ecosystems employ sandboxing mechanisms?  Are certain operating systems offering a greater proportion of sandboxed software than others, and how do the functionalities of these sandboxes differ? In pursuit of answers, we examined source package archives for four open source OSs (Table~\ref{tab:os-versions}) to identify packages that utilize sandboxing mechanisms.

\begin{table}
  \centering
  \caption{Different OSs examined and their packages.}
  \begin{tabular}{llr}
    \toprule
    OS & Version & Packages \\
    \midrule
    Debian Linux & Bullseye 11 & 59,913\\
    Fedora Linux & 37 & 66,166 \\
    OpenBSD & 7.3 & 7,787\\
    FreeBSD & 13.2 & 30,766 \\
    \bottomrule
  \end{tabular}
  \label{tab:os-versions}
\end{table}

\vspace\baselineskip\noindent
Specifically, this paper addresses the following research questions:
\begin{itemize}
    \item RQ1: To what extent sandboxing mechanisms are used in operating systems packages?
    \item RQ2: How do sandboxing mechanisms compare in terms of complexity and security when applied to the same package in different operating systems?
\end{itemize}

We observed that although many packages rely on packages that employ sandboxing mechanisms, the actual number of packages directly using sandboxing APIs appears relatively small across all operating systems. When examining the packages that utilize these mechanisms, we found instances where developers didn't fully use the capabilities of fine-grained sandboxing mechanisms, such as those in Seccomp. Instead, they tended to simplify their approaches, mapping it to simpler models like Pledge's.

Additionally, we found that using sandboxing mechanisms requires developers to restructure their code to align with sandboxing requirements. Interestingly, Pledge's model of sandboxing sometimes offers features that result in a more restrictive sandbox compared to Seccomp and Capsicum.

\section{Background}

\subsection{Sandboxing mechanisms}
\subsubsection{Seccomp}

\emph{Seccomp} was introduced in Linux in 2005 (short for secure computing), offering developers an API to limit the system calls their applications can make. It is inspired by the now-defunct \emph{Systrace} \cite{kurchuk2004recursive}. With \emph{Seccomp}, developers can create allow or deny list of system calls and configure eight different actions for filtered calls. For example, \emph{Seccomp} can trap or kill a process upon execution of a system call, preventing attackers from relaunching their attacks. \emph{Seccomp} Filters are inherited by child processes/threads.

Developers using \emph{Seccomp} can write Berkeley Packet Filter (BPF) code to specify allowed system calls and their arguments. BPF code, resembling assembly language, runs in the kernel and has restrictions, such as a maximum size limit and a limited set of instructions \cite{calavera2019linux}. The BPF interface enables examination of system call arguments, but it can't dereference pointer arguments. This limitation is to avoid Time-of-Check to Time-of-Use (TOCTOU) attacks \cite{canella2021domain}. For instance, while you can create BPF filters allowing the \texttt{read} system call with specific file descriptors, filtering the \texttt{fopen} system call and checking its string path argument isn't possible since the argument is a pointer.

\emph{Seccomp} provides a C library (libseccomp) that enables filtering system calls without the need to write BPF code directly. Listing ~\ref{lst:seccomp_example} shows an example of \emph{Seccomp} used with libseccomp, restricting a process to solely execute the \texttt{write} system call when the passed argument corresponds to the stdout file descriptor.

\begin{lstlisting}[caption={Seccomp filter that allows the write system call to stdout file descriptor},label={lst:seccomp_example},language=C,breaklines,captionpos=b]
scmp_filter_ctx ctx;
ctx = seccomp_init(SCMP_ACT_KILL);      
seccomp_rule_add(ctx, SCMP_ACT_ALLOW, SCMP_SYS(write), 1, SCMP_A0(SCMP_CMP_EQ, STDOUT_FILENO));
seccomp_rule_add(ctx, SCMP_ACT_ALLOW, SCMP_SYS(exit_group), 0);
seccomp_load(ctx);
\end{lstlisting}    

\subsubsection{Landlock}
\emph{Landlock} was introduced as a Linux security module in 2017 \cite{salaun2017landlock}. It empowers developers to associate access rights rules with files and directories, enabling, for example, restrictions like read-only access to the filesystem. \emph{Landlock} offers four access rights for files and eleven for directories. While it allows additional filters to restrict process access, it doesn't permit their expansion. Notably, files or directories opened before creating the sandbox remain unrestricted, and similar to \emph{Seccomp}, child processes inherit the sandbox from their parent. The example in Listing ~\ref{lst:landlock_example} demonstrates \emph{Landlock} usage to confine process access to read-only permissions on the /usr directory, extracted and simplified from \emph{Landlock} man pages.

\begin{lstlisting}[caption={Landlock used to only allow read access to /usr folder},label={lst:landlock_example}, captionpos=b]
landlock_create_ruleset(&attr, sizeof(attr), 0);
struct landlock_path_beneath_attr path_beneath = {0};
path_beneath.allowed_access = LANDLOCK_ACCESS_FS_EXECUTE | LANDLOCK_ACCESS_FS_READ_FILE | LANDLOCK_ACCESS_FS_READ_DIR;
path_beneath.parent_fd = open("/usr", O_PATH | O_CLOEXEC);
landlock_add_rule(ruleset_fd, LANDLOCK_RULE_PATH_BENEATH, &path_beneath, 0);
close(path_beneath.parent_fd);
prctl(PR_SET_NO_NEW_PRIVS,1, 0, 0, 0)
landlock_restrict_self(ruleset_fd, 0)
\end{lstlisting}

\subsubsection{Pledge and Unveil}
OpenBSD provides \emph{Pledge} and \emph{Unveil} system calls, which are similar to \emph{Seccomp} and \emph{Landlock} but have a simpler interface and user experience \cite{pledgeunveil}. Previously known as \emph{Tame}, \emph{Pledge} allows developers to declare that their program will only use specific promises. Promises are collections of privileges and system calls with similar functionality, defining the scope of permissible actions. For example, the \texttt{unix} promise groups 10 system calls related to operating on the \texttt{AF\_UNIX} domain, while the \texttt{id} promise encompasses system calls related to process rights and identity \cite{pledge}.

\emph{Unveil} limits applications to only use specified directories with designated permissions, though not as finely detailed as those provided by \emph{Landlock}. Developers can employ \emph{Unveil} with permissions like \texttt{r}, \texttt{w}, \texttt{x}, \texttt{c}, representing read, write, execute, and create/remove permissions for the path \cite{unveil}.

Any attempt to use permissions outside the pledged promises will result in the program being terminated, unless the \texttt{error} promise is used, in which case an error is returned. Unlike \emph{Seccomp}, child processes don't inherit the sandbox restrictions.

\begin{lstlisting}[caption={Example of Pledge and Unveil},language=C,breaklines,captionpos=b]
pledge("stdio tty", NULL);
unveil("/folder",   "r");
\end{lstlisting}

\subsubsection{Capsicum}
In FreeBSD, \emph{Capsicum} serves as a sandboxing mechanism designed to help developers restrict their applications' resource access \cite{zaborski2016capsicum}. \emph{Capsicum} equips developers with various system calls to control their application's access to file descriptors, covering files, sockets, and other entities represented as file descriptors in the kernel \cite{watson2010capsicum}. Once in the capability mode, achieved through \verb|cap_enter|, the process is barred from accessing the global namespace, including reading from the filesystem. Additionally, \emph{Capsicum} enables the assignment of capability rights and permissions to file descriptors. For example, this can be used to permit only reading access to a specific file. Moreover, \emph{Capsicum} facilitates the management of allowed IOCTLs and FCNTLs numbers for an application. Accessing resources beyond the allowed scope doesn't terminate the program but triggers a capability violation error. Similar to \emph{Seccomp}, child processes of a capability-mode process will also be sandboxed.

The following example, extracted from the FreeBSD \emph{Capsicum} man pages, illustrates how \emph{Capsicum} can be employed to restrict a file descriptor to read-only permission.

\begin{lstlisting}[caption={Capsicum code that restricts a file descriptor to read-only mode },captionpos=b]
cap_rights_t setrights;
char buf[1];
int fd;
fd = open("/tmp/foo", O_RDWR);
if (fd < 0)
    err(1, "open() failed");
if (cap_enter() < 0)
    err(1, "cap_enter() failed");
cap_rights_init(&setrights, CAP_READ);
if (cap_rights_limit(fd, &setrights) < 0)
    err(1, "cap_rights_limit()	failed");
buf[0] = 'X';
if (write(fd, buf,sizeof(buf)) > 0)
    errx(1, "write() succeeded!");
if (read(fd, buf, sizeof(buf)) < 0)
    err(1, "read() failed");
\end{lstlisting}

\section{Method}

To assess the adoption of sandboxing mechanisms in software across different ecosystems, we downloaded the source code for all packages within the x86-64 Debian and Fedora repositories, as well as the ports trees of OpenBSD and FreeBSD. This data collection and analysis took place from April to August 2023. We specifically chose these operating systems because they represent two major Linux distributions, forming the foundation for numerous other distributions that build upon their software. Additionally, FreeBSD and OpenBSD, though less prevalent, offer distinctive sandboxing mechanisms and cater to specific niches within the broader open source OS ecosystem. OpenBSD, in particular, is known for prioritizing security and actively implementing sandboxing measures for applications, making it a noteworthy focus for investigation.

To identify packages that use sandboxing mechanisms, we used regular expressions to search code that invokes sandbox mechanisms system calls. Specifically, we searched within \texttt{.c}, \texttt{.cpp}, \texttt{.cc}, \texttt{.h}, \texttt{.hpp}, and \texttt{.sh} files for calls to:
\begin{itemize}
\item \texttt{seccomp\_init}
\item \texttt{landlock\_restrict\_self}
\item \texttt{pledge}
\item \texttt{unveil}
\item \texttt{cap\_enter}
\item \texttt{cap\_rights\_limits}
\item \texttt{cap\_ioctl\_limits}
\item \texttt{cap\_fcntls\_limit}
\end{itemize}
For \emph{Seccomp} and \emph{Landlock}, we also searched for invocations of \texttt{prctl} and \texttt{syscall} with the appropriate arguments.

We took measures to minimize false positives, such as filtering out noise like system call invocations in commented code, documentation, or strings. We also filtered out instances where sandboxing mechanisms are employed solely in test cases rather than for creating a sandbox. We observed a number of compiler packages containing \emph{llvm} source code where \emph{Seccomp} filters are installed only to test whether ASAN (Address Sanitizer) works properly under sandbox restrictions. Similarly, various kernel-related packages include tests to ensure that \emph{Seccomp} and \emph{Landlock} function correctly. However, none of these packages utilize sandboxing mechanisms to establish a sandbox; the mechanisms are used only in testing code. We also manually checked packages that passed these filters to make sure that we didn't include any false positives. 

Additionally, our analysis accounts for instances where multiple packages share the same source code. For example, packages like \emph{firefox-esr-l10n-en-gb} and \emph{firefox-esr-l10n-en-ca} in Debian share identical source code despite generating different packages (distinguished by language). Our analysis considers results from all packages, even if they originate from the same source code.

We also compared how different operating systems make use of their sandboxes. We focused on packages existing in different operating systems that utilize at least two different sandboxing mechanisms in their codebases. Our selection aimed to highlight interesting usage patterns and compare how developers implement different sandboxing mechanisms for the same package across different operating systems. This delves into the security implications and complexity associated with implementing sandboxes.

\section{Threats to validity}

There are several threats to validity of this study, the largest of which relates to how we identify sandboxed packages.
We focus our analysis on C and C++ source code since not all sandboxing mechanisms have official libraries for the other languages, but all make system calls that are exposed through C and C++ bindings. To help account for this when looking at whether a package makes use of sandboxing features we first look for packages directly calling the APIs, and also for packages declaring a dependency on packages which make use of a sandboxing API.
This means that if a package provides native bindings for a sandbox API (such as \texttt{ruby-pledge} in OpenBSD which provides Ruby bindings for the \texttt{pledge} syscall) then we capture programs that use sandboxing features through the API but without making C or C++ syscalls directly.  It also means that if a program depends on a sandboxed binary for some tiny aspect of its functionality then we treat the whole package as being sandboxed, as in \texttt{pssh}, \texttt{ssvnc}, \texttt{vagrant} and many other packages that depend on \texttt{openssh-client} package which is sandboxed using Seccomp. Whilst in truth it may be only some minor aspect that is palmed off on an external library.  We favor including these packages in our analysis as it's not our place to determine to what extent a package is sandboxed or how appropriately; but it is a weakness of the study.

Sometimes the source code of a package is unavailable due to unmaintained websites or other problems. We failed to download the source code of less than 4\% of the packages for any OS and didn't include these packages in our study.  Additionally, for the BSDs core system software (such as the default shell) is distributed as part of the OS release and not as part of an individual package.  This software is also omitted.

Another weakness is that whilst packages are \emph{predominantly} used for packaging software, sometimes they're not; some packages just add documentation, for example.
Our analysis doesn't exclude non-software packages since it is challenging to identify them.

\section{Results}

\subsection{Sandboxing adoption}

\begin{table}
  \centering
  \caption{Percentage of packages that use sandboxing mechanisms.}
  \begin{tabular}{cccc}
  \toprule
  OS &
Packages count &
Sandboxed &
Depend on Sandboxed \\
  \midrule
  Debian &
59893 &
406 (0.68\%) &
10164 (16.97\%) \\
  Fedora &
63706 &
396 (0.62\%) &
23747 (37.28\%) \\
  OpenBSD &
7700 &
114 (1.48\%) &
1903 (24.71\%) \\
  FreeBSD &
30763 &
43 (0.14\%) &
13900 (45.18\%) \\
  \bottomrule
\end{tabular}

  \label{table:per}
\end{table}  

\begin{table}
  \centering
  \caption{Number of packages in each OS with sandboxing mechanisms, per sandboxing tool.} 
  \begin{tabular}{ccccc}
  \toprule
  OS &
Seccomp &
Pledge &
Unveil &
Capsicum \\
  \midrule
  Debian &
406 &
x &
x &
x \\
  Fedora &
396 &
x &
x &
x \\
  OpenBSD &
x &
111 \footnotemark &
36 &
x \\
  FreeBSD &
x &
x &
x &
43 \\
  \bottomrule
\end{tabular}

  \label{table:quan}
\end{table}

Table \ref{table:per} shows the percentage of packages that use sandboxing mechanisms in the studied OSs and table \ref{table:quan} shows usage statistics for each of the sandboxing mechanisms. Whilst each OS contains thousands of packages, only a fraction of them (for all OSs <1\%) ever seem to use the sandbox APIs directly. Instead, packages seem to \emph{depend} on other packages (that is, required and installed first) that either use or again depend on sandboxed code. \footnotetext{There are 3 packages that use Unveil without Pledge.}

\emph{Landlock} is only used in kernel packages in testing code and a sample program, but nothing aside from that, it's not used in any other packages to create a sandbox.

\subsection{What kinds of apps get sandboxed?}

We categorized packages that directly use sandboxing mechanisms, each operating system has its own category system but to allow comparison between OSs we used Debian's label for other OSs. Table ~\ref{table:cats} shows the 10 most popular categories in packages that use sandboxing mechanisms in all operating systems.

For Linux, the majority of the packages that make use of sandboxed code are for development and libraries (Table~\ref{table:cats}), but for OpenBSD and FreeBSD it is instead the networking and utility applications that mostly make use of sandboxing mechanisms. Looking at results without context can be misleading, for instance, Debian has 104 browser packages, this is because 97 of them are different versions of \emph{firefox}, while Fedora has 4, and FreeBSD and OpenBSD have 2 packages for \emph{firefox}. When taking a closer look at the development packages in both Fedora and Debian, we notice that they are mostly packages for popular hypervisors like \emph{zen} and \emph{qemu} and containers solutions like \emph{lxc}, there are also packages for fuzzing and testing tools like \emph{rumur} and \emph{afl}. In OpenBSD, the development category contains a more diverse set of software such as development frameworks \emph{arcan}, web applications \emph{sblg}, \emph{stagit} and development tools such as \emph{openradtool} and \emph{the\_silver\_searcher}. 

Network tools are prime candidates for sandboxing as they are typically exposed to external users. The BSDs have a higher percentage of packages under the networking category that employ a sandboxing mechanism. Most of the networking packages in Debian and Fedora are different versions of \emph{ntp}, \emph{openssh} and \emph{gnutls}, for instance, Fedora has 10 \emph{gnutls} packages and Debian has 5 \emph{openssh} packages. The BSDs, especially OpenBSD has a broader and more diverse software under the networking category, such as IRC clients \emph{catgirl}, Gemini \emph{gmid} and Gopher \emph{geomyidae} servers, and VPN software like \emph{mlvpn}.

Overall there is a diverse range of apps that make use of sandboxing mechanisms from document readers, administrator tools, and graphics tools to games, such as \emph{warzone2100} that uses \emph{Pledge}, and \emph{redeclipse} that uses \emph{Seccomp}. Mostly, however, the mechanisms are used in networking and development packages.

\begin{table*}[t]
  \centering
  \caption{Categories of packages that directly use sandboxing mechanisms. The table includes the 10 most used categories across all OSs.}
  \begin{tabular}{ccccccccccc}
  \toprule
  OS &
devel &
libs &
net &
browser &
utils &
admin &
mail &
javascript &
misc &
python \\
  \midrule
  Debian &
34 (8\%) &
105 (26\%) &
28 (7\%) &
104 (26\%) &
11 (3\%) &
36 (9\%) &
67 (17\%) &
1 (0\%) &
10 (2\%) &
3 (1\%) \\
  Fedora &
141 (36\%) &
62 (16\%) &
41 (10\%) &
13 (3\%) &
40 (10\%) &
55 (14\%) &
3 (1\%) &
28 (7\%) &
7 (2\%) &
4 (1\%) \\
  OpenBSD &
11 (10\%) &
2 (2\%) &
38 (33\%) &
3 (3\%) &
40 (35\%) &
4 (4\%) &
4 (4\%) &
1 (1\%) &
1 (1\%) &
0 \\
  FreeBSD &
2 (5\%) &
3 (7\%) &
12 (28\%) &
3 (7\%) &
8 (19\%) &
5 (12\%) &
2 (5\%) &
2 (5\%) &
0 &
1 (2\%) \\
  \bottomrule
\end{tabular}

  \label{table:cats}
\end{table*}

\subsection{What is the sandboxed code doing?}

\subsubsection{Linux}

With \emph{Seccomp}, developers can examine system calls' arguments and limit their program to run system calls only if certain arguments are provided. For instance, you can limit programs to only write to \texttt{stdout} or limit the \texttt{ioclt} system call to be run only with specific numbers. To do that, developers have to write BPF filters. BPF filters have a syntax similar to assembly and can be challenging to write. One would assume that developers would tend to simplify their \emph{Seccomp} sandbox and only use it to allow/deny system calls without going the extra mile and write BPF code to limit their arguments. Our analysis shows the opposite. We manually checked packages that use \emph{Seccomp} and found that 79\% of them in Fedora and 82\% in Debian utilize it to check system calls' arguments and further restrict their sandbox. When checking for packages that use libseccomp, we found that 40\% of packages in Debian and 76\% in Fedora use libseccomp and the rest write BPF code to implement the \emph{Seccomp} sandbox.

\subsubsection{OpenBSD}

We collected promises that are used with every call to \emph{Pledge}. The \texttt{stdio} promise is the most used promise, it is used in almost every call to \emph{Pledge}. The \texttt{stdio} promise grants access to 69 system calls that are mostly related to working with standard input and output and are essential for libc to work. Followed by \texttt{rpath} which allows system calls that have a read-only effect, \texttt{wpath} that allows system calls that have a write-only effect, and \texttt{cpath} that allows system calls that create new files or directories \cite{pledge}. Software, in general, uses the file system in one way or another, and these results suggest that a sandboxing mechanism that offers the ability to restrict the file system actions will be appealing to and used by developers, in contrast to some promises that are never used in any of the packages in OpenBSD such as \texttt{mcast} that allows operating on multicast sockets, \texttt{tape} which allows certain IOCTLs on tape drivers, \texttt{settime} that allows setting the system time and \texttt{error} which makes \emph{Pledge} returns an error instead of killing the program upon a violation of the pledged promises.

\subsubsection{FreeBSD}
\emph{Capsicum} is used by 43 packages in the FreeBSD ports tree. Capsicum provides different system calls to enable sandboxing, the \texttt{cap\_enter} system call that puts the process in capability mode restricting access to the global namespaces is used in 34 packages, while \texttt{cap\_rights\_limit} is used 27 times, this system call is used to impose restrictions on file descriptors. The \texttt{cap\_ioctls\_limit} system call controls the access to IOCTLs and is used 7 times. The \texttt{cap\_fcntls\_limit} that manages access to \texttt{fcntl} commands is used 6 times.

\subsection{What are they sandboxing?}

Are sandboxes for similar packages equivalent between OSs?  
We searched the database for packages that exist in different operating systems and use their sandboxing mechanisms, then we examined how different sandboxes were applied to secure the packages across different operating systems. We were interested in observing how developers utilize the sandboxing mechanisms when used to secure the same package in terms of the security and complexity of the implemented sandbox.  Whilst this isn't an exhaustive comparison, we can survey how some packages differ when run on different OSs.

Having a fine-grained sandboxing mechanism like \emph{Seccomp} enables developers to create detailed sandboxes that only allow what an application needs and nothing more, in contrast to sandboxing mechanisms that don't allow that level of detailed sandboxing like \emph{Pledge}. This was the case for a number of applications that we examined, like \emph{arping} package that exists in Linux, OpenBSD, and FreeBSD. It uses \emph{Pledge} with only two promises, \texttt{stdio}, and \texttt{tty}. The \texttt{stdio} promise allows 69 system calls that allow the program to perform basic I/O operations using the standard input and output, and it allows the functionality of libc to work. The \texttt{tty} promise grants the application read and write permissions on a \texttt{/dev/tty} device, and it grants permissions to perform 11 IOCTL operations that are related to controlling and interacting with the terminal device. The package uses \emph{Unveil} to grant access to all of the root filesystem. However, \emph{arping} package implements a stricter and more secure sandbox using \emph{Seccomp}. It limits the process access to only 9 system calls, \texttt{fstat}, \texttt{write}, \texttt{ioctl}, \texttt{sendto}, \texttt{select}, \texttt{pselect6}, \texttt{newfstatat}, \texttt{exit\_group} and \texttt{rt\_sigreturn}.  In addition, the sandbox utilizes BPF filters to limit \texttt{fstat} and \texttt{write} system calls to use only standard output and error and to limit the numbers passed to both the \texttt{ioctl} and \texttt{sendto} system calls. In FreeBSD, the package doesn't use any sandboxing mechanisms.

But fine-grained sandboxing doesn't always seem to be used to create a stricter and more detailed sandboxing. The \emph{xwallpaper} package exists in Debian, FreeBSD and OpenBSD, and it uses \emph{Pledge} and \emph{Seccomp} to create sandboxes. Contrary to \emph{arping} package, the sandbox created by \emph{Seccomp} isn't more secure or restricted than the \emph{Pledge} sandbox. The package uses \emph{Pledge} twice, at the beginning of the main function, \emph{Pledge} is used with 6 promises, \texttt{dns}, \texttt{inet}, \texttt{proc} \texttt{rpath} \texttt{stdio} and \texttt{unix}, and later, \emph{Pledge} is used to drop privileges and only allow one promise, \texttt{stdio}. When using \emph{Seccomp}, the package has two functions to implement the \emph{Seccomp} sandbox, \texttt{stage1\_sandbox()} and \texttt{stage2\_sandbox()}. In both functions, the developer clearly attempted to mimic the sandbox created by \emph{Pledge}. The developer used \emph{Seccomp} to create allow list of system calls that match exactly what \emph{Pledge} promises map to, this is shown also in the code comments as shown below. In `stage2\_sandbox' function, the developer allow-listed the 69 system calls that are granted by the stdio promise.

\begin{lstlisting}[caption={Using Seccomp to mimic Pledge},captionpos=b]
/* pledge: dns */
seccomp_rule_add(ctx, SCMP_ACT_ALLOW, SCMP_SYS(connect), 0)
seccomp_rule_add(ctx, SCMP_ACT_ALLOW, SCMP_SYS(sendto), 0)
seccomp_rule_add(ctx, SCMP_ACT_ALLOW, SCMP_SYS(socket), 0)
...
/* pledge: inet+unix */
seccomp_rule_add(ctx, SCMP_ACT_ALLOW, SCMP_SYS(accept), 0)
seccomp_rule_add(ctx, SCMP_ACT_ALLOW, SCMP_SYS(accept4), 0)
...
\end{lstlisting}

When compared to \emph{Pledge}, \emph{Seccomp} and \emph{Capsicum} seem to be more flexible and allow for detailed implementation of sandboxes. However, this isn't always the case, \emph{Pledge} has some features that allow developers to design more secure sandboxes. This is apparent when used in the \emph{gmid} package, the developer of this package wrote a blog about his experience when adopting the different sandboxing mechanisms to \emph{gmid} package \cite{Omar_Polo_2021}. 

The \emph{gmid} package is a Gemini server that listens on a port, serves static files, and runs CGI scripts. To accommodate the sandboxing mechanisms and to implement a secure sandbox, the developer had to split the application into two processes, one that listens on the port and the other one to execute CGI scripts on demand. This split is necessary because child processes inherit the sandbox of the parent process in both \emph{Seccomp} and \emph{Capsicum}, and without the split, it would not be possible to fork and run CGI scripts since they would be running under the parent sandbox and might not execute correctly since they might need more permissions than the parent has. By splitting the program into two processes, the listener and the executor, it would be possible to use \emph{Seccomp} and \emph{Capsicum} to sandbox the listener process but not the executor process. However, with \emph{Pledge} and \emph{Unveil}, it was possible to sandbox both processes since the sandbox isn't inherited by the child process when forked to execute CGI scripts. Despite \emph{Pledge} and \emph{Unveil} being the least fine-grained when compared to other sandboxing mechanisms, this flexibility allows for creating a more secure sandbox than other mechanisms. 

Another case is \emph{lowdown}, which is a library that uses \emph{Pledge} and \emph{Capsicum} to create a sandbox. Examining the source code reveals that the developers had intentions to adopt \emph{Seccomp} but didn't since there are checks in the config files and tests for \emph{Seccomp}, but it's not used in the actual library code. \emph{Pledge} is used in two functions, \texttt{sandbox\_pre} and \texttt{sandbox\_post} with \texttt{stdio}, \texttt{rpath}, \texttt{wpath}, and \texttt{cpath} promises in the \texttt{sandbox\_pre} and only \texttt{stdio} promise in \texttt{sandbox\_post}. With \emph{Capsicum}, only the \texttt{sandbox\_post} function is implemented and \emph{Capsicum} is used to limit the process access to \texttt{stdin}, \texttt{stdout}, and \texttt{stderr} with respective permissions and then entering the capability mode with \texttt{cap\_enter}. The developer was able to sandbox the program in its early stages using \emph{Pledge} and specify system calls that are needed at the start of the program and then drop privileges to only allow the \texttt{stdio} promise but didn't do that with \emph{Capsicum}.

\section{Discussion}

The results shed light on the adoption of sandboxing mechanisms and the way developers use these mechanisms indicates that there is a space for improvement and that the adoption of sandboxing mechanisms can be further increased. 

\subsection{Sandboxed from the off}

Examining the post-boot processes in the different operating systems shows intriguing findings. As depicted in Table \ref{table:past_boot}, OpenBSD stands out with 36 processes running after boot, of which 33 employ either \emph{Pledge} or \emph{Unveil}. Following that, Fedora has 16 processes out of 138 utilizing \emph{Seccomp}, while Debian has 5 out of 120 processes that use \emph{Seccomp}. With FreeBSD, only 3 out of 47 processes utilize \emph{Capsicum}. Although OpenBSD exhibits a significant advantage over other operating systems in this context, this distinction is less pronounced when considering software packages.

Whilst the number of packages that directly invoke sandboxing APIs seems to be very low in all operating systems (OpenBSD has the highest percentage), the number of packages that depend on them is much higher. Understanding whether an app is appropriately sandboxed becomes difficult when it's not immediately apparent how a sandbox is deployed.  For example, is it accurate to label \texttt{deepin-terminal} as sandboxed simply because it relies on a chain of dependencies, eventually reaching \texttt{openssh-client}, which is sandboxed using \emph{Seccomp}? While one might argue that this minimal dependency on \texttt{openssh-client} brings a level of sandboxing for a specific aspect, the broader application may still be vulnerable. 
 
Despite these considerations, a more thorough understanding of these sandboxing mechanisms can contribute to their improved adoption. Security software comes with challenges that might affect its adoption \cite{dewitt2007usability}, and we believe there is a need to better understand the space of problems presented by these mechanisms. But what are these problem(s)?

\begin{table}
  \centering
  \caption{Processes that make use of sandboxing features immediately on boot following a clean install.}
    \begin{tabular}{ccc}
          \toprule
          OS &  Post-boot processes & Sandboxed \\
          \midrule
          Debian & 120 & 5 (4.16\%)\\
          Fedora & 137 & 16 (11.67\%)  \\
          OpenBSD & 36 & 33 (91.66\%) \\
          FreeBSD & 47 & 3 (6.38\%) \\
          \bottomrule \\
      \end{tabular}
    \label{table:past_boot}
\end{table}

\subsection{Uncharted territory}

One part of the problem appears to be the lack of understanding of the developers' mindset and rationale. What do developers look for in a sandboxing mechanism? Are developers facing usability issues with the current ones? Do developers use these mechanisms correctly?

Our results show that \emph{Pledge} and \emph{Unveil}, despite being developed recently, have a higher usage percentage than \emph{Seccomp} in Debian or Fedora packages. \emph{Pledge} and \emph{Unveil} have a very simple interface and are easy to integrate with programs. There are efforts to port \emph{Pledge} and \emph{Unveil} to Linux by open-source developers and this demonstrates an interest in using these mechanisms. 

When looking at \emph{Seccomp} and \emph{Landlock}, such mechanisms allow implementing fine-grained sandboxes, but they need more work from the developers to apply them when compared to \emph{Pledge} and \emph{Unveil}. During our analysis, we noticed that developers write a lot more code to adopt \emph{Seccomp} when compared to other sandboxing mechanisms. For instance, gmid package has 549 lines of code to implement \emph{Seccomp} sandboxing while only have 28 lines to adopt \emph{Capsicum} and 96 to adopt \emph{Pledge} and \emph{Unveil}. We also noticed that a lot of packages write hundreds of lines of codes as wrappers for \emph{Seccomp} API as in \emph{brltty} and \emph{crun} packages.

We noticed that sometimes developers copy \emph{Seccomp} filters code from other packages as in \emph{ansilove}, \emph{bdf2sf} and \emph{logswan} that share similar code of \emph{Seccomp} filters. In addition, some packages share a list of system calls that are considered dangerous and should be denylisted, we noticed this in \emph{flatpak} package where developers stated in the code comments that the \emph{Seccomp} denylist has been copied from \emph{linux-user-chroot} which is influenced by \emph{sandstorm.io} denylist. While in these cases, we didn't notice any issues with the shared code, this behavior is concerning as research shows that developers tend to copy code without giving it any security considerations ~\cite{fischer2017stack}. 

So is the problem that some sandboxing mechanisms are complex and not easy to use? One would assume that when using \emph{Seccomp}, developers will tend to avoid the complexities of writing BPF filters, but we noticed a lot of packages use \emph{Seccomp} with BPF filters to scan system calls arguments. BPF filters are difficult to write, yet developers are willing to go the extra mile and implement a more secure sandbox even if it requires complex implementations.

All of above begs the question of whether we really need more fine-grained sandboxing or should operating systems maintainers focus on creating simple sandboxing mechanisms. There is a general lack of understanding of what developers need in a sandbox and the struggles they face when using the current mechanisms. Understanding these areas will provide insights into how the adoption of sandboxing mechanisms can be increased.

\subsection{It is not easy}

Still, even with simple mechanisms like \emph{Pledge} and \emph{Unveil}, the adoption levels seem to be very low, there have to be other challenges developers are facing. One problem is that in order for developers to use any of the studied mechanisms, they have to have a solid understanding of what the application is doing, what system calls it is using, and which files it needs access to. This becomes more challenging when the program to be sandboxed is large and complex. In addition, as seen in the cases we presented earlier, sometimes developers have to re-architecture their applications and split them into smaller units and processes to accommodate sandboxing. This burden can be one of the reasons that developers steer away from adopting such technologies.

Furthermore, debugging \emph{Seccomp} and \emph{Pledge} is challenging. When using a functionality outside the promised ones, the program will exit, and produce a core dump, and the developer has to either analyze the core dump or debug the program to figure out the violated promise. With \emph{Seccomp}, the developer has to use external tools like \emph{strace} or write debugging logic to be able to capture what went wrong. 

\subsection{But how can we fix this?}

Research shows that developers prioritize functionality over security, and they show indifference towards security \cite{arizon2021understanding, dewitt2006aligning}. One solution to this problem is to have programming languages and platforms to implement security measures by default. One example of this is Deno, a Javascript runtime environment that is used to build web applications.  By default, Deno doesn't grant applications any access to system resources, and the developer is required to explicitly ask for permissions to the resources using an API ~\cite{denopermissions}. Such an approach doesn't leave an option for developers but to adopt security. Furthermore, having tools in operating systems that allow developers to check if their application is over-privileged can guide developers to utilize these mechanisms better.

\section{Related work}

\subsection{Adoption of security mechanisms}

There is no research that focuses on the adoption of sandboxing solutions, but some research investigated the adoption challenges of security tools in general from both the social and the technical aspects. In their research, Witschey et al. quantified developers' adoption of security tools by conducting an online survey with developers to determine the factors that could determine security tools adoption \cite{witschey2015quantifying}. They found out that social and cultural factors such as interaction with the security community and the reputation of the security tool represent the most important factors that affect security tool adoption. In addition, their research highlights many innovative factors that affect the adoption of security tools. One of these factors is how policy-makers might be able to increase security tool adoption by increasing observability. This means that developers are more likely to use a security tool when they observe other developers using it. This can be done with sandboxing by showing that certain packages use sandboxing mechanisms to maximize the security of their application. Another factor is the complexity, their survey found that complexity is usually linked to the ability of the security tool and that security tool designers can increase adoption by reducing the complexity of security tools. 

Linden et al{.}'s research examines the rationale behind developers' decisions when it comes to decisions that affect the security of the application \cite{van2020schrodinger}. They conducted two studies and found that developers might be writing secure code but not secure apps. Their research provided insights that help in understanding developers' rationale and they provided suggestions to help developers make security-aware decisions, they suggested enhancing development IDEs with features that incorporate support and help in making the right choices when it comes to security features and mechanisms.

\subsection{Research on sandboxing mechanisms}

Previous research examined the sandboxing space and categorized sandboxing solutions based on many factors. Maass et al. conducted a systematic literature review of sandboxing solutions \cite{maass2016systematic}. Their research discusses many definitions of the sandbox term and that it could be an isolation or policy enforcement technique. The research discussed the lack of validation techniques for present solutions and that most of the current solutions focus on the performance overhead and they ignore validating other important aspects such as the usability of the sandbox. The research provides recommendations to better validate sandboxes such as using assurance case diagrams. In addition, the paper discusses the lack of research on the usability of sandboxing and provides insights into how the usability of sandboxes can be assessed.

Schreuders et al. conducted a survey of sandboxing solutions focusing on application-oriented solutions \cite{schreuders2013state}. They categorized application-oriented solutions into rule-based and isolation based and went further to discuss the limitations and challenges of the adoption of these sandboxes. The research outlines that the challenges of the adoption of studied sandboxes stem from their usability challenges and policy complexity. 

\subsection{Usability of security mechanisms}

The research highlighted the importance of having usable security APIs and that usability should be considered well when designing security software ~\cite{yee2004aligning,green2016developers}. Research indicates a strong correlation between security and usability, emphasizing that disregarding usability can have dire consequences~\cite{acar2017comparing}. However, many security software suffers from usability issues, and developers struggle to use security APIs and software. Schreuders et al.
examined the usability of three access-control mechanisms, including SELinux, AppArmor, and FBAC-LSM by conducting a usability study ~\cite{schreuders2012towards}. Their results illustrated many usability issues with SELinux and AppAromor and they proposed changes to these mechanisms to avoid the found usability issues. DeWitt and Kuljis conducted a usability study on Polaris (a security-related software that is designed to be usable) \cite{dewitt2006aligning}. Their research found that
developers showed indifference towards security and that despite Polaris software being designed to be usable, it still had many usability issues. Furthermore, their research found that aligning usability with security becomes more challenging when the usability aspect is considered later in the development process. Patnaik~et~al{.} research examined the usability of cryptography APIs by looking at 2000 StackOverFlow questions which studied whether cryptography libraries violate Green~and~Smith
principles. They identified many usability smells in cryptography APIs and grouped them into categories-\emph{whiffs} \cite{patnaik2019usability}.

\section{Conclusion}

While sandboxing mechanisms play a pivotal role in enhancing security, our findings reveal a nuanced landscape. Many packages incorporate sandboxing mechanisms indirectly, yet only a limited number directly use these mechanisms. The diversity of sandboxing mechanisms presents both opportunities and challenges for developers aiming to implement secure software across multiple platforms. The cases we presented suggest that, despite their apparent simplicity, mechanisms like \emph{Pledge} and \emph{Unveil} are capable of implementing strict sandboxes when compared to other fine-grained mechanisms. Furthermore, our research highlights the lack of understanding of the challenges that developers face when using these mechanisms and whether these mechanisms are designed to fit developers' needs in the first place. Addressing these questions will provide further insights into how to drive developers to use sandboxing mechanisms more and to better understand the lack of adoption of these mechanisms in open-source ecosystems.

\footnotesize \bibliographystyle{ACM-Reference-Format}
\bibliography{paper.bib}


\end{document}